\documentclass{article}

\usepackage{PRIMEarxiv}

\usepackage[utf8]{inputenc} 
\usepackage[T1]{fontenc}    
\usepackage{hyperref}       
\usepackage{url}            
\usepackage{booktabs}       
\usepackage{amsfonts}       
\usepackage{nicefrac}       
\usepackage{microtype}      
\usepackage{lipsum}
\usepackage{fancyhdr}       
\usepackage{graphicx}       
\usepackage{subfig}
\graphicspath{{media/}}     

\title{TranSimHub: A Unified Air–Ground Simulation Platform for Multi-Modal Perception and Decision-Making}

\author{%
    Maonan Wang$^{1,2}$ \quad 
    Yirong Chen$^{2,3}$ \quad
    Yuxin Cai$^4$ \quad
    Aoyu Pang$^1$ \quad
    Yuejiao Xie$^1$ \quad
    Zian Ma$^5$ \\
    \textbf{Chengcheng Xu}$^5$ \quad
    \textbf{Kemou Jiang}$^6$ \quad
    \textbf{Ding Wang}$^2$ \quad
    \textbf{Laurent Roullet}$^7$ \quad
    \textbf{Chung Shue Chen}$^7$ \\
    \textbf{Zhiyong Cui}$^6$ \quad
    \textbf{Yuheng Kan}$^5$ \quad
    \textbf{Michael Lepech}$^3$ \quad
    \textbf{Man-On Pun}$^1$ \\
    $^1$The Chinese University of Hong Kong, Shenzhen \quad 
    $^2$Shanghai AI Laboratory \quad
    $^3$Stanford University \\
    $^4$Nanyang Technological University \quad 
    $^5$SenseTime Group Ltd \quad
    $^6$Beihang University \quad
    $^7$Nokia Bell Labs \\
}

\begin{document}
\maketitle

\vspace{-20pt}
\begin{figure}[!h]
  \centering
  \includegraphics[width=\linewidth]{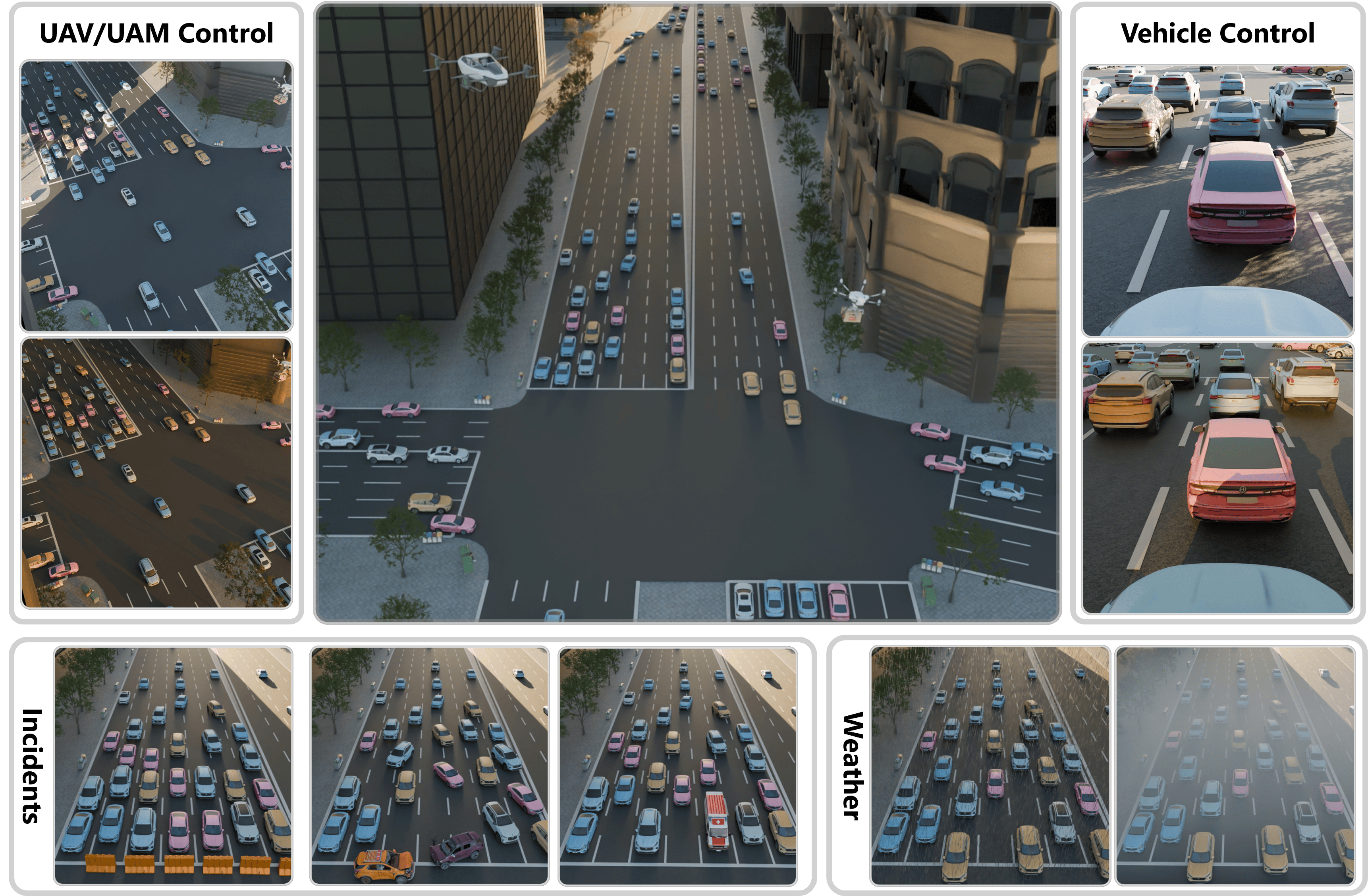}
  \caption{Example scene generated by TranSimHub. The central panel illustrates the integrated air–ground environment, where ground entities coordinate with aerial systems such as UAVs and UAMs. The side panels show multi-perspective rendering and control capacities for both aerial and ground agents at different times of day, including noon and dusk. The lower row demonstrates the platform’s causal scene editing capabilities, such as inserting traffic accidents, deploying emergency vehicles, and modifying environmental conditions like rain and fog.}
  \label{fig:overall}
\end{figure}

\begin{abstract}
Air–ground collaborative intelligence is becoming a key approach for next-generation urban intelligent transportation management, where aerial and ground systems work together on perception, communication, and decision-making. However, the lack of a unified multi-modal simulation environment has limited progress in studying cross-domain perception, coordination under communication constraints, and joint decision optimization. To address this gap, we present TranSimHub, a unified simulation platform for air–ground collaborative intelligence. TranSimHub offers synchronized multi-view rendering across RGB, depth, and semantic segmentation modalities, ensuring consistent perception between aerial and ground viewpoints. It also supports information exchange between the two domains and includes a causal scene editor that enables controllable scenario creation and counterfactual analysis under diverse conditions such as different weather, emergency events, and dynamic obstacles. We release TranSimHub as an open-source platform that supports end-to-end research on perception, fusion, and control across realistic air and ground traffic scenes. Our code is available at \url{https://github.com/Traffic-Alpha/TransSimHub}. 
\end{abstract}


\section{Introduction}


The rapid development of intelligent transportation systems (ITS) and urban air mobility (UAM) is transforming the landscape of modern cities. With the emergence of autonomous vehicles, unmanned aerial vehicles (UAVs), aerial taxis (UAMs), and intelligent infrastructures, modern mobility systems are evolving toward air–ground collaborative intelligence, where aerial and ground entities jointly perceive, communicate, and make decisions. Such collaboration enables integrated situational awareness, efficient traffic flow, and improved safety across heterogeneous transportation layers, forming a key foundation for next-generation intelligent urban management.

Although significant progress has been made in both ground-based and aerial simulation research, existing platforms remain isolated and domain-specific. Ground simulators such as VISSIM \cite{fellendorf2010microscopic}, Aimsun \cite{casas2010traffic}, and SUMO \cite{SUMO2018} have advanced studies in autonomous driving and traffic signal control, yet they focus mainly on road-level interactions among vehicles, traffic lights, and pedestrians, without incorporating aerial perception or communication. Conversely, aerial simulators such as AirSim \cite{shah2017airsim} $Fe^{3}$ \cite{xue2018fe3} and PX4 \cite{meier2015px4} specialize in flight control, path planning, and visual perception for UAVs or UAMs but lack modeling of ground dynamics or infrastructure coordination. Consequently, researchers currently lack a unified framework to investigate multi-agent cooperation, cross-domain perception fusion, and decision-making under communication constraints—capabilities essential for large-scale, intelligent air–ground mobility systems.

To address these limitations, we present TranSimHub, an open-source simulation platform that unifies air–ground collaborative intelligence within a single, extensible environment. TranSimHub models diverse aerial entities (e.g., UAVs, UAMs) and ground entities (e.g., vehicles, pedestrians, and traffic infrastructures) in a shared, dynamically configurable 3D world. The platform supports synchronized multi-modal rendering—including RGB, depth, and semantic segmentation—from multiple viewpoints such as drone cameras, intersection cameras, and vehicle-mounted sensors, enabling integrated research on perception, communication, and decision coupling. Furthermore, TranSimHub incorporates a causal scene editor that allows users to manipulate environmental conditions, insert special events (e.g., accidents, emergency vehicles), and generate counterfactual scenarios to evaluate model robustness, causal generalization, and safety-critical behaviors.

We envision TranSimHub as a foundation for end-to-end research on perception, fusion, and control across realistic air–ground traffic environments. By offering standardized interfaces and modular design, the platform enables researchers to focus on algorithmic innovation while ensuring reproducibility and interoperability across heterogeneous transportation domains.


\section{System Overview of TranSimHub}

\begin{figure}[!h]
  \centering
  \includegraphics[width=0.8\linewidth]{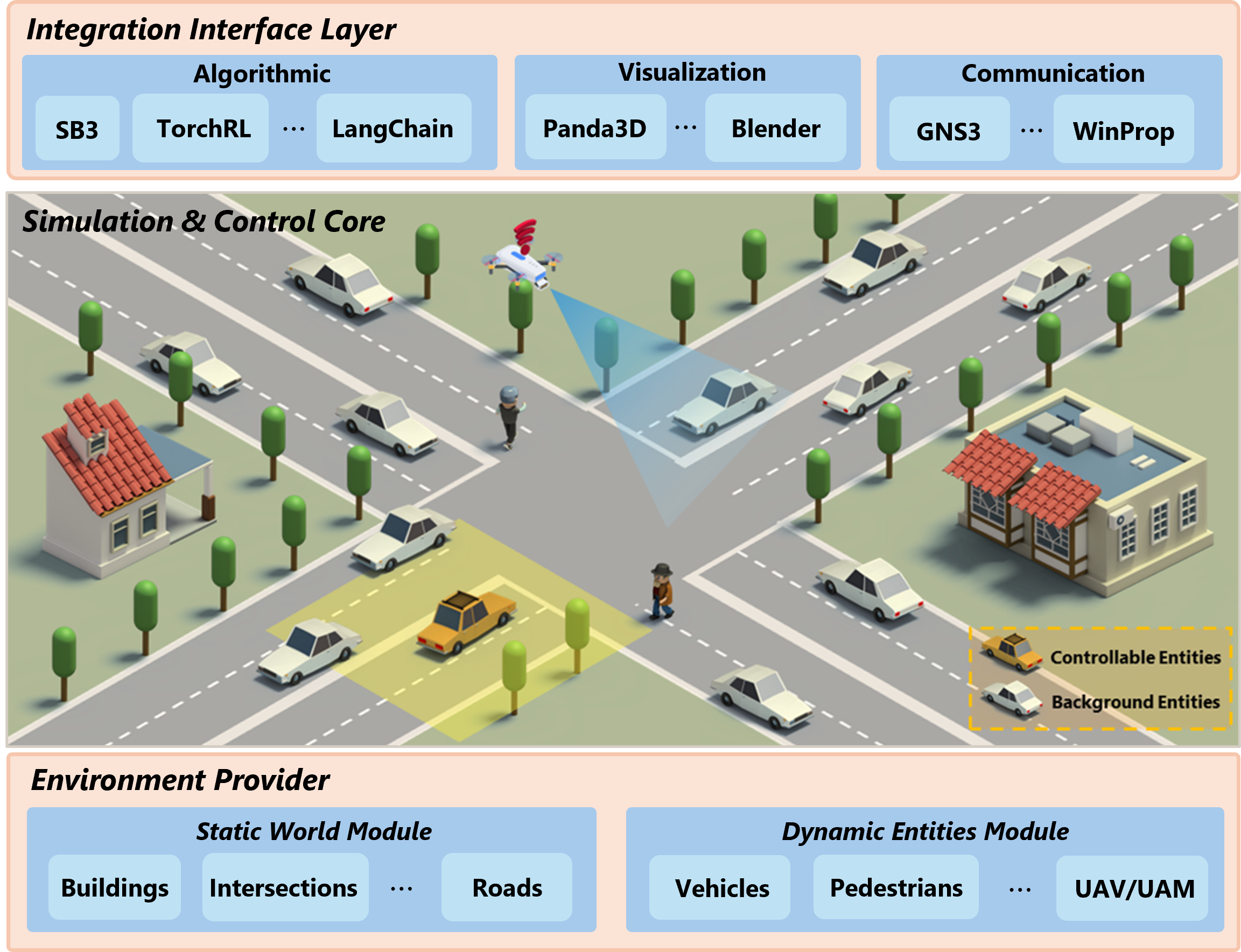}
  \caption{TranSimHub Architecture. It comprises the Environment Provider Layer (static and dynamic entities), the Simulation \& Control Layer (unified APIs and scene interaction), and the Integration Interface Layer (connections to algorithmic, visualization, and communication modules).}
  \label{fig:framework}
\end{figure}

TranSimHub is structured into three layers, as shown in Fig.~\ref{fig:framework}. The \textbf{Environment Provider Layer} offers both static and dynamic components of the simulated world. Static elements—such as buildings, intersections, and roads—are imported from OpenStreetMap (OSM) and can be further customized to define map geometry, lane attributes, and signal configurations. Dynamic entities include vehicles, pedestrians, and UAVs, which are controlled through predefined engines such as SUMO, or alternatively by user-defined strategies. Both ground and aerial agents support policy-level customization, allowing integration with RL- or LLM-driven control frameworks. To support large-scale simulations, all modules are designed to be hot-swappable, enabling selective loading of components and efficient computation.

The \textbf{Simulation and Control Layer} forms the core of TranSimHub, managing interactions and control logic among entities. Entities are categorized as controllable or background depending on their accessibility. Users can specify control policies for controllable entities, while background entities follow predefined behaviors from environment providers. The simulator allows the perception range of controllable entities to be customized, limiting their observations to nearby agents and thereby improving computational efficiency.

The \textbf{Integration Interface Layer} bridges TranSimHub with external ecosystems through standardized and extensible APIs. A Gym-compatible interface ensures seamless interoperability with reinforcement learning frameworks such as Stable Baselines3 \cite{raffin2021stable} and TorchRL \cite{bou2023torchrl}, supporting both single-agent and multi-agent learning paradigms. (M)LLM-driven control is enabled through LangChain \cite{Chase_LangChain_2022}, facilitating natural language–based interaction with simulated agents. For high-fidelity visualization, TranSimHub leverages Blender \cite{Blender2022} for more realistic rendering and also supports Panda3D \cite{goslin2004panda3d} for high efficiency simulation. In addition, GNS3 \cite{neumann2015book} and WinProp \cite{hoppe2017wave} are integrated to model inter-entity communication and wireless propagation, enabling comprehensive studies on perception, coordination, and decision-making across air–ground collaborative systems.

\subsection{Multi-Modal Multi-View Rendering}

A core feature of TranSimHub lies in its ability to produce synchronized, high-fidelity visual outputs from multiple viewpoints and modalities, enabling comprehensive studies on perception, fusion, and decision-making across aerial and ground domains. The rendering engine is built upon the Blender framework and allows cameras to be mounted on various entities, including vehicles, drones, and infrastructure, ensuring simultaneous rendering from heterogeneous perspectives within a shared simulation timeline.

\begin{figure*}[!ht]
    \centering
    \subfloat[]{\includegraphics[width=\textwidth]{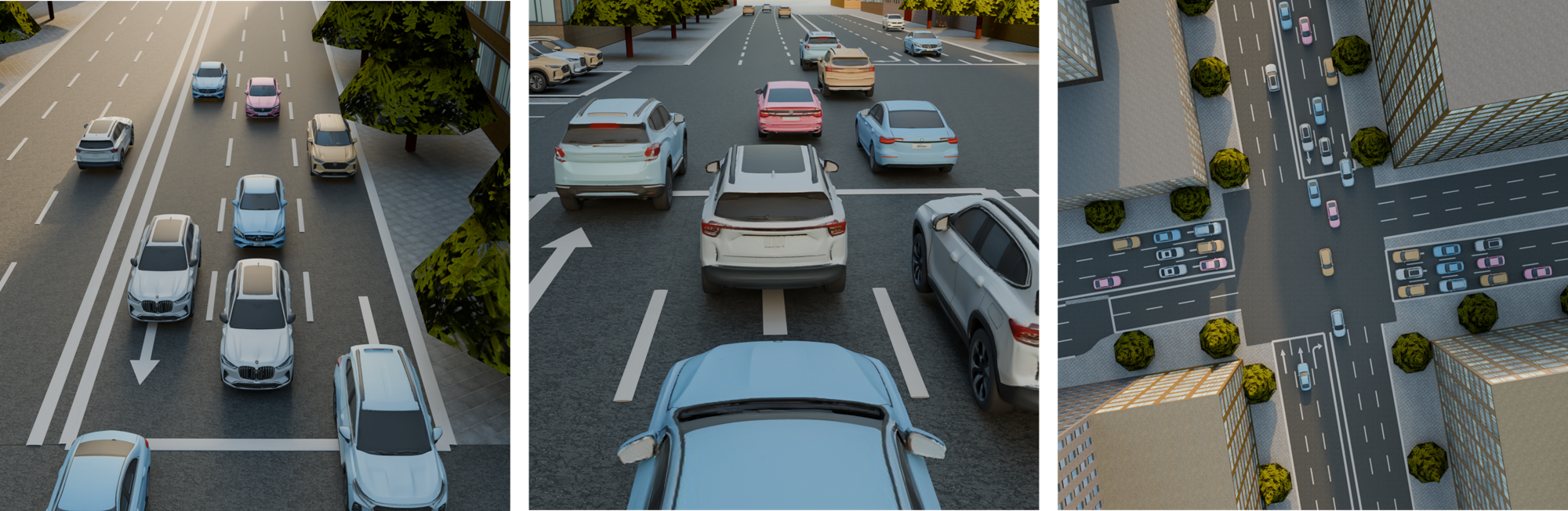}
    \label{fig:views}}
    
    \hfill
    
    \subfloat[]{\includegraphics[width=\textwidth]{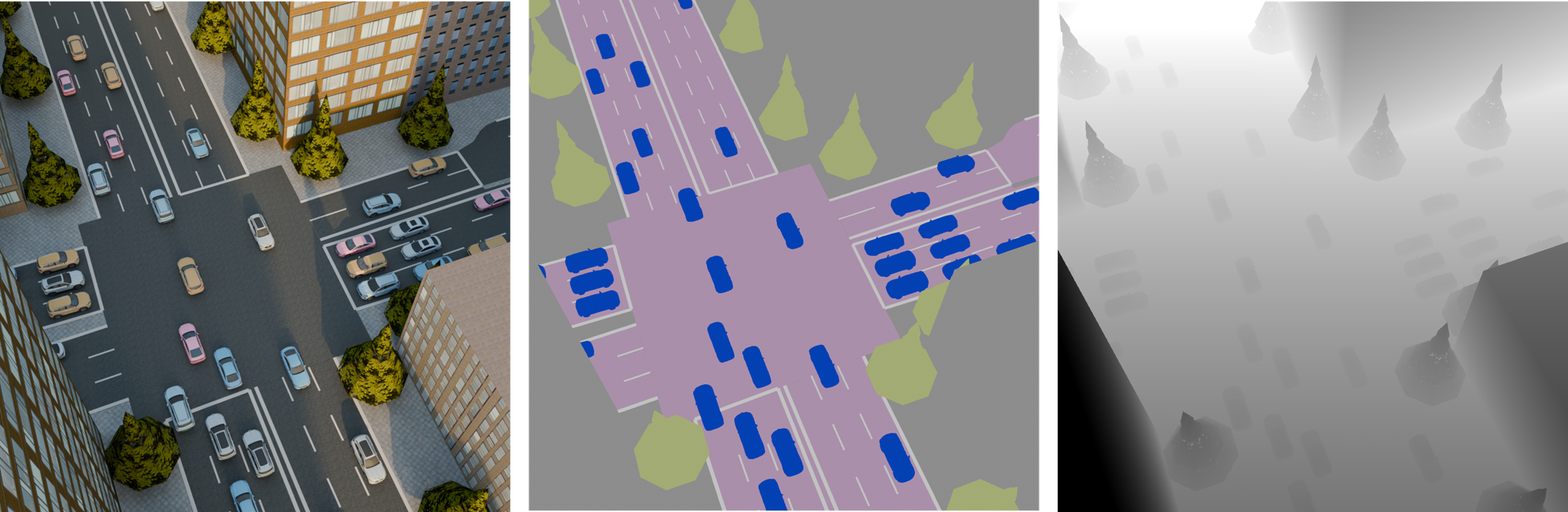}
    \label{fig:output_formats}}
    \caption{Multi-modal and multi-view rendering in TranSimHub. (a) Examples of different camera perspectives supported by the platform, including an intersection-level view, a first-person view from a selected vehicle, and an aerial top-down view from a UAV or UAM. (b) Examples of multi-modal rendering outputs from the same scene, including RGB, semantic segmentation, and depth maps, all synchronized in both space and time.}
    \label{fig:multi_views}
\end{figure*}

As illustrated in Fig.~\ref{fig:views}, TranSimHub supports diverse camera perspectives that reflect the heterogeneity of real-world deployment settings. The platform provides fixed intersection-level viewpoints for capturing traffic flow and pedestrian interactions, vehicle-mounted first-person viewpoints for studying autonomous driving and lane perception, and aerial viewpoints from UAVs or UAMs for monitoring large-scale traffic dynamics. These perspectives can operate concurrently within a single simulation, and all camera poses can be conveniently specified through configuration files. This flexibility enables researchers to explore cooperative perception and sensor fusion across agents operating at different altitudes, positions, and orientations.

Beyond spatial diversity, TranSimHub enables multi-modal rendering to accommodate a wide range of perception and learning tasks. As shown in Fig.~\ref{fig:output_formats}, each viewpoint can generate synchronized outputs in RGB, semantic segmentation, and depth formats. The RGB modality provides a photorealistic appearance suitable for visual perception and imitation learning, whereas the semantic segmentation maps deliver pixel-level understanding of scene components such as vehicles, pedestrians, and road markings. Depth maps further enrich these representations by encoding geometric and distance information essential for spatial reasoning and 3D reconstruction. All modalities are temporally and spatially aligned, allowing precise cross-modal correspondence within the same frame. The rendering pipeline supports real-time frame export and flexible format conversion, facilitating seamless integration with downstream machine learning frameworks for both training and evaluation. In addition, users can selectively render one or multiple modalities to balance data richness and computational efficiency.

Through this unified design, TranSimHub establishes a synchronized and extensible environment for generating multi-view and multi-modal data. This capability provides a powerful foundation for investigating cooperative perception, cross-view fusion, and air–ground collaborative decision-making in complex and dynamic urban environments.

\begin{figure*}[!ht]
    \centering
    \subfloat[]{\includegraphics[width=\textwidth]{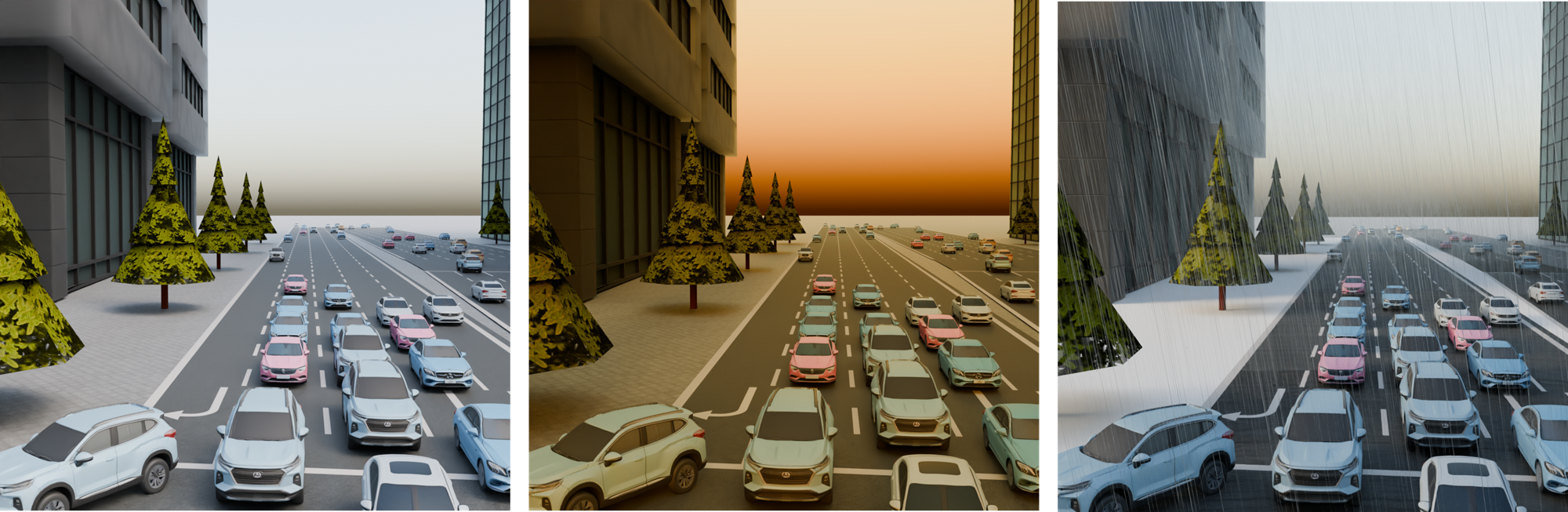}
    \label{fig:weather}}
    
    \hfill
    
    \subfloat[]{\includegraphics[width=\textwidth]{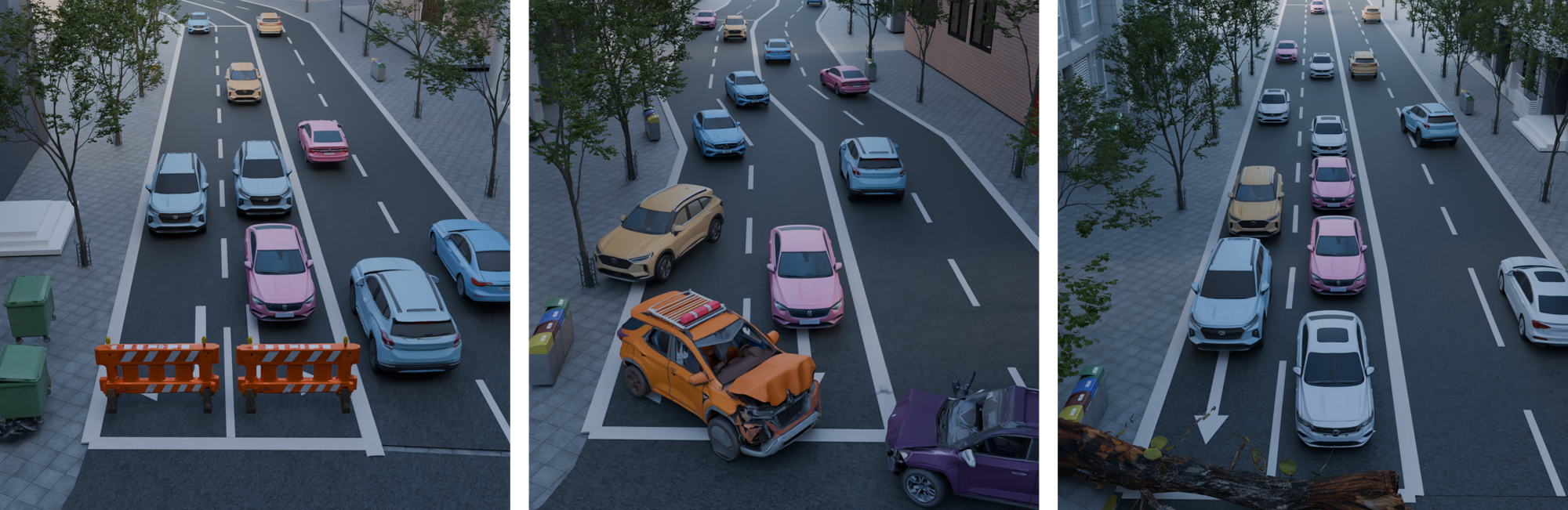}
    \label{fig:dynamic_events}}

    \hfill

    \subfloat[]{\includegraphics[width=\textwidth]{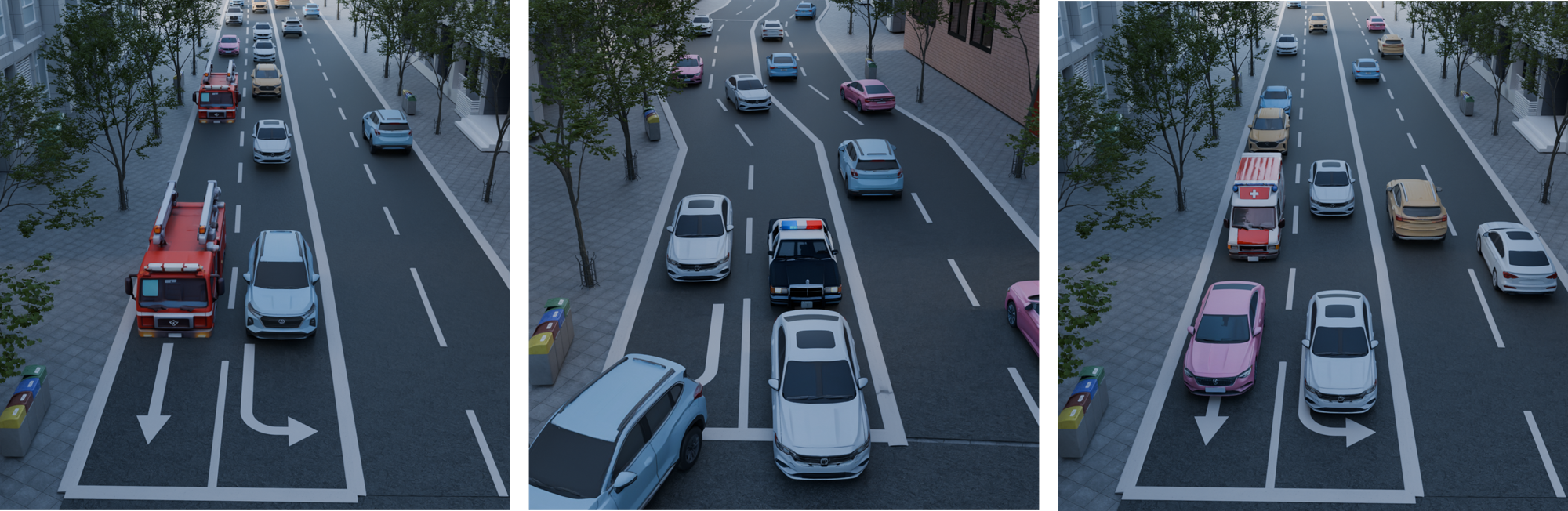}
    \label{fig:emergency_vehicles}}
    
    \caption{Causal scene editing in TranSimHub. (a) Examples of different weather and lighting conditions, including daytime, dusk, and rain; (b) Examples of dynamic events such as temporary road closures with barriers, traffic collisions, and fallen trees; (c) Examples of special-purpose vehicles including fire trucks, police cars, and ambulances.}
    \label{fig:scene_editing}
\end{figure*}

\subsection{Causal Scene Editing}

Beyond static environment construction, TranSimHub provides a causal scene editing module that allows users to manipulate environmental factors and dynamic events to create diverse and controllable simulation scenarios. This module serves as an essential tool for evaluating system robustness, safety, and generalization under variable and unexpected conditions. All editing operations are implemented through a config file, ensuring reproducibility and consistency across experiments.

As illustrated in Fig.~\ref{fig:weather}, TranSimHub supports the modification of weather and time conditions, enabling scenes to be rendered under different illumination and atmospheric settings such as clear daytime, cloudy dusk, and rainfall. These variations allow perception and control models to be evaluated across diverse visual domains and sensor degradations caused by lighting changes or weather interference. 

In addition to environmental settings, TranSimHub enables the injection of dynamic events that simulate unexpected or safety-critical situations in real traffic environments. As shown in Fig.~\ref{fig:dynamic_events}, the platform supports several representative events, including temporary road closures using barriers, traffic accidents involving collisions, and fallen trees caused by strong winds or typhoons. Each event can be precisely positioned and parameterized, allowing researchers to study how perception and decision models respond to abnormal conditions and occlusions. The causal manipulation of these events facilitates counterfactual experiments—where a single environmental factor is altered while others remain constant—making it possible to analyze causal relationships between scene changes and model behaviors.

Furthermore, TranSimHub incorporates special-purpose emergency vehicles, such as fire trucks, police cars, and ambulances, as depicted in Fig.~\ref{fig:emergency_vehicles}. These vehicles can be introduced into the scene with distinct visual appearances and dynamic trajectories. This feature enables simulation of emergency scenarios, right-of-way coordination, and traffic control interactions, providing a realistic testbed for algorithms dealing with multi-agent decision-making and cooperative response mechanisms.

Through these capabilities, TranSimHub establishes a flexible and causally controllable simulation environment. The integration of weather variations, dynamic events, and emergency vehicles enables researchers to conduct systematic analyses of model robustness, interpretability, and causal generalization, advancing the study of perception–communication–control coupling in complex and safety-critical air–ground scenarios.

\section{Related Work}

\subsection{Ground Simulators}

Ground-based traffic simulators have long served as the cornerstone of urban mobility research, offering quantitative tools to model vehicle interactions, multi-modal flows, and network-level dynamics. Early traffic assignment systems such as DynaMIT \cite{ben2002real} and DynaSmart \cite{mahmassani2001dynamic} enabled the analysis of dynamic routing and congestion propagation under varying demand conditions, laying the conceptual groundwork for subsequent microscopic and mesoscopic simulation frameworks. However, these early systems were often constrained by limited flexibility in behavioral modeling and scalability across city-scale networks.

Building upon these foundations, modern simulators such as VISSIM \cite{fellendorf2010microscopic}, Aimsun \cite{casas2010traffic}, and SUMO \cite{SUMO2018} have become the standards for high-fidelity ground traffic modeling. VISSIM and Aimsun emphasize detailed car-following and lane-changing mechanisms while supporting corridor- to region-scale simulations with customizable signal control logic. In contrast, SUMO, as a fully open-source platform, prioritizes extensibility and reproducibility, facilitating integration with reinforcement learning and optimization pipelines for adaptive traffic management. Together, these simulators provide versatile foundations for analyzing congestion, traffic signal coordination, and multi-modal interactions within complex urban environments.

As simulation requirements have expanded from flow-level dynamics to human-centric behavior, the focus has increasingly shifted toward agent-based and activity-based frameworks such as MATSim \cite{w2016multi}, BEAM \cite{gopal2017modeling}, and POLARIS \cite{auld2016polaris}, which simulate individual decision-making, mode choice, and temporal scheduling at large scales. These models bridge microscopic realism with macroscopic scalability, enabling system-level analyses of transport policy, energy consumption, and social equity.

Meanwhile, specialized simulators have emerged to support distinct research domains. CARLA \cite{dosovitskiy2017carla}, MetaDrive \cite{li2022metadrive}, SMARTS \cite{pmlr-v155-zhou21a} and Flow \cite{wu2021flow} provide photorealistic and physics-based environments for autonomous driving and control research, while CityFlow \cite{zhang2019cityflow}, CityFlowER \cite{da2024cityflower} and Libsignal \cite{mei2024libsignal} focus specifically on large-scale traffic signal optimization with efficient parallel architectures. 

\subsection{Air Simulators}

Airspace simulation platforms have evolved in parallel with ground-based systems, forming the analytical foundation for research on air traffic management and UAM. Early simulators such as FACET \cite{bilimoria2001facet} and ATM TestBed \cite{robinson2017development}, developed by NASA, focused on trajectory-level modeling of aircraft operations within the National Airspace System (NAS), enabling large-scale evaluations of traffic flow management, sector capacity, and conflict detection. However, these systems were primarily designed for conventional aviation and thus lack the scalability and flexibility required to simulate dense, low-altitude UAM environments. To address these limitations, a new generation of airspace simulators has been developed with an emphasis on flexibility, scalability, and modularity. Fe³ \cite{xue2018fe3} provides GPU-accelerated simulations for high-density low-altitude flight scenarios, supporting rapid evaluation of separation assurance and contingency management strategies. VertiSim \cite{yedavalli2021simuam} models vertiport-level operations such as layout design, pad scheduling, and resource allocation, offering a practical environment for assessing takeoff and landing throughput constraints. Meanwhile, DTALite \cite{zhou2014dtalite} and the open-source BlueSky \cite{hoekstra2016bluesky} extend air traffic simulation to metropolitan and national scales, enabling trajectory optimization and concept-of-operations validation for emerging UAM networks. Collectively, these simulators represent a shift from traditional air traffic modeling toward integrated, multi-scale frameworks that can accommodate future UAM operations.

Beyond operational modeling, several simulators have been designed to support environmental, policy, and perception-oriented research, further broadening the scope of airspace simulation. AEDT \cite{roof2007aviation} and AEIC \cite{simone2013rapid} integrate noise, emission, and performance models to quantify the ecological footprint of aviation activities, while AirTraf 2.0 \cite{yamashita2020newly} couples flight trajectory simulation with the EMAC atmospheric chemistry model to study contrail formation and climate effects. In parallel, new simulators such as OpenUAV \cite{gao2025openfly} and OpenFly \cite{wang2024towards} extend airspace simulation into the vision-language navigation domain, allowing aerial agents to perform perception-grounded reasoning and goal-directed navigation in photorealistic 3D environments. Together, these platforms form a layered and interdisciplinary simulation ecosystem—from vertiport micro-operations to macro-scale environmental analyses—laying the groundwork for integrating multimodal perception, language understanding, and decision-making into next-generation UAM and air–ground collaborative systems.

Despite their sophistication, existing air simulators generally operate in isolation from terrestrial mobility systems, lacking unified temporal, spatial, and semantic synchronization with ground-level simulators. This fragmentation limits the ability to study air–ground interactions, such as integrated scheduling, multi-modal transfer, and shared infrastructure utilization. To bridge this gap, TranSimHub is designed as a unified air–ground simulation platform that synchronizes aerial and ground perspectives, supports cross-domain information exchange, and enables controllable scenario generation for comprehensive evaluation of air–ground collaborative intelligence.

\section{Conclusion}

In this work, we presented \textbf{TranSimHub}, a unified simulation platform for air–ground collaborative intelligence that integrates aerial and ground domains within a configurable 3D environment. The platform supports synchronized multi-modal rendering, cross-domain communication, and causal scene editing, enabling joint research on perception, coordination, and decision-making across heterogeneous agents. By providing standardized interfaces compatible with reinforcement learning, language-driven control, and communication modeling, TranSimHub facilitates end-to-end experimentation on perception fusion and policy learning. We envision it as a foundation for next-generation urban intelligence research, promoting reproducibility, interoperability, and cross-disciplinary innovation in realistic air–ground mobility systems.

\bibliographystyle{unsrt}  
\bibliography{references}

\end{document}